\documentclass[proceedings]{JHEP3} 

\conference{International Workshop on Astroparticle and High Energy
Physics}

\usepackage{epsfig,multicol}			

\newbox\mybox
\newcommand\fverb{\setbox\mybox=\hbox\bgroup\verb}
\newcommand\fverbdo{\egroup\medskip\noindent\fbox{\unhbox\mybox}\ }
\newcommand\fverbit{\egroup\item[\fbox{\unhbox\mybox}]}

\title{Warm inflation solution to the eta-problem}

\author{
Arjun Berera
	\thanks{Speaker, PPARC Advanced Fellow} \\
	School of Physics, University of Edinburgh, Scotland, EH9 3JZ
\\ E-mail: \email{ab@ph.ed.ac.uk}
}

\conference{AHEP-2003}

\abstract{ 
Warm inflationary dynamics is shown to satisfy both the slow-roll
and density perturbation constraints for $m_{\phi} \gg H$
or equivalently $\eta \gg 1$ and for inflaton field amplitudes
much below the Planck scale, $\langle \phi \rangle < m_{pl}$.  
I start by reviewing 
the two types of inflation dynamics, isentropic 
or cold inflation and nonisentropic or warm inflation. 
In the former, inflation occurs without radiation production, 
whereas in the latter both radiation production and 
inflation occur concurrently.  I then discuss recent, detailed, quantum field
theory calculations showing that many generic inflation
models, including hybrid inflation, which were believed only
to have cold inflation regimes, in fact have regimes of both warm
and cold inflation.  These results dispel many foregone assumptions
generally made up to now about inflation models and bring to
the fore various elementary issues that must be addressed to
do reliable calculations from inflation models.
I also discuss density perturbations and observational consequences of warm 
inflation, especially related to WMAP.  Finally 
I show that warm inflation has intrinsic features 
that make the ``eta problem'' nonexistent, and field
amplitudes are below the Planck scale. 
}

\begin{document}

\section{Introduction}
\label{sect1}

The main requirement of particle physics, arising from the density perturbation
and slow-roll conditions of inflation, is a very flat potential.
The degree of flatness necessary is typically expressed through the
slow-roll parameter 
\begin{equation}
\eta \equiv \frac{m^2_{pl}}{8 \pi V} \left(\frac{d^2V}{d\varphi^2}\right).
\label{etadef}
\end{equation}
In standard inflation models
\cite{Guth:1980zm, Sato:ds,Albrecht:1982wi,Linde:1981mu},
where inflaton evolution is damped
by the term $3H {\dot \varphi}$, the slow-roll condition
amounts to $\eta \stackrel{<}{\sim} 1$, which equivalently means
the potential can not have mass terms bigger than $\sim H^2 \varphi^2$.

Since Supersymmetry suppresses quantum corrections, thus can
preserve the tree level potential, it has been a central idea
in realizing such flat inflationary potentials.  Of course, since inflation
requires a nonzero vacuum energy density, inevitably
SUSY must be broken during the inflation period,
thus possibly ruining the desired degree of flatness in the potential.
In particular, once supergravity effects are included, it
becomes very difficult for this symmetry to preserve flatness
at the leevl of $\eta < 1$.  For F-term inflation, where the nonzero
vacuum energy density arises from terms in the superpotential,
no symmetry prohibits the appearance of the
Planck mass suppressed higher dimensional operators
$a_n \varphi^n/m_{pl}^{n-4}$
\cite{Copeland:1994vg,Gaillard:1995az,Kolda:1998kc,Arkani-Hamed:2003mz} 
For the large class of chaotic inflation
type models \cite{Linde:gd}, where inflation occurs with the inflaton
field amplitude above $m_{pl}$, to control these higher dimensional
operators would require the fine-tuning of a infinite number
of parameters.  Even for models where inflation occurs
for field amplitudes below the Planck scale, dimension six operator
terms of the form $V \varphi^2/m^2_{pl} \sim H^2 \varphi^2$ can
emerge and ruin the desired flatness.  Both minimal and nonminimal
Kahler potentials can lead to such terms \cite{Arkani-Hamed:2003mz}.

One possible solution to the $\eta$-problem might be D-term inflation.
In such models, the nonzero vacuum energy arises from the
supersymmetrization of the gauge kinetic energy.
However a closer examination 
\cite{Kolda:1998kc,Arkani-Hamed:2003mz} 
reveals that
attaining the required degree of flatness makes such models
very restrictive.

Up to now, attempts to solve the eta-problem have sought
symmetries that can maintain this desired degree of flatness.
One of the few that has proven successful is called the
Heisenberg symmetry \cite{Gaillard:1995az}, although it is very restrictive. 
Another proposal has been a certain shift 
symmetry \cite{Arkani-Hamed:2003mz},
which is particularly interesting as it does not require SUSY.
In common, all attempts so far have one foregone conclusion,
that inflaton dynamics is only viable for $\eta < 1$.
However, if the inflaton evolution happened to have a damping term
larger than $3H{\dot\varphi}$, then clearly slow-roll
can be satisfied for $\eta > 1$. Such a possibility is
precisely what occurs in warm inflationary dynamics.

There are two distinct dynamical realizations of inflation.
In the original picture, termed cold, supercooled or isentropic
inflation
\cite{Guth:1980zm,Sato:ds,Albrecht:1982wi,Linde:1981mu,Linde:gd},
the universe rapidly supercools during inflation and
subsequently a reheating phase is invoked to end inflation and
put the universe back into a radition dominated regime.
In the other picture, termed warm or nonisentropic inflation
\cite{Berera:1995ie},
dissipative effects are important during the inflation period,
so that radition production occurs concurrently with inflationary 
expansion.
Phenomenologically, the inflaton evolution in simple warm inflation models
has the form,
\begin{equation}
{\ddot \varphi} + [3H+\Upsilon(\varphi)] {\dot \varphi} +
\xi {\rm R} \varphi + \frac{dV_{\rm eff}(\varphi)}{d\varphi} = 0 ,
\label{amapprox}
\end{equation}
where 
$R = 6({\ddot a}/{a} + {\dot a^2}/{a^2})$
is the curvature scalar.
For $\Upsilon =0 $, this is the familiar inflaton evolution equation
for cold inflation, but for a nonzero $\Upsilon$, it corresponds to
the case where the inflaton field is dissipating energy
into the universe, thus
creating a radiation component.

Despite a historical belief that cold inflation is 
the most common form of inflationary dynamics, recent
work has shown that warm inflationary dynamics also is very
generic \cite{Berera:2001gs}.  
As will be discussed later in the talk, many simple models
which up to now have been believed to exclusively yield
cold inflation, in fact also have warm inflationary regimes.
Very elementary thermodynamic considerations already 
point to the fact that
cold inflation is a very restrictive picture.
In particular, even if
the inflaton dissipated a minuscule fraction
of its energy, say one part in $10^{20}$,
it still would constitute a significant radiation
energy density component in the universe.
{}For example, for inflation with vacuum (i.e. potential) energy
at the GUT scale $\sim 10^{15-16} {\rm GeV}$,
leaking one part in $10^{20}$ of this energy density into radiation 
corresponds to a temperature of $10^{11} {\rm GeV}$,
which is nonnegligible.  In fact, the most relevant
lower bound that cosmology places on the temperature after inflation
comes from the success of hot Big-Bang nucleosynthesis,
which requires the universe to be within the
radiation dominated regime by $T \stackrel{>}{\sim} 1 {\rm GeV}$.
This limit can be met in the above example by dissipating 
as little as one part in $10^{60}$ of the vacuum energy
density into radiation.
Thus, from the perspective of both interacting field theory and
basic notions of equipartition,  it appears to be a highly
tuned requirement of cold inflation to prohibit
the inflaton from such tiny amounts of dissipation.

In this talk, I review the progress made in developing
warm inflation.  In Sec. \ref{sect2} recent work
on quantum field theory first principles
calculations which realize the inflaton effective 
equation of motion (EOM) Eq. (\ref{amapprox})
is given.  In particular, a very simple model
involving just four fields is shown to be adequate for
realizing warm inflation.  In Sec. \ref{sect3}, I review the
theory of density perturbations in warm inflation.
In this picture, density perturbations are thermally induced and
formulas for inflaton fluctuations are given.
I then discuss recent work which has numerically evolved the
cosmological perturbation equations under warm inflationary conditions.
The process of dissipation during warm inflation leads to many interesting
features in the scalar spectral index.  One noteworthy result,
which I focus on, is for a simple symmetry breaking potential,
which leads to a spectral index that is blue at large scales
and red at small scales, similar to the spectrum suggested by
recent WMAP results \cite{Bennett:2003bz,Peiris:2003ff}.
Finally in Sec. \ref{sect4}, I show how warm inflation dynamics
based on Eq. (\ref{amapprox}) and with the thermal inflaton fluctuations of
Sec. \ref{sect3}, has no "eta-problem" and has inflaton field amplitudes
much below the Planck scale.

\section{Quantum field theory dynamics}
\label{sect2}

Our previous considerations of inflaton dynamics were limited to
nonexpanding spacetime.
In addition the earliest of these works looked for high temperature
warm inflation solutions, under rigid adiabatic, equilibrium
conditions \cite{Berera:1998gx}.  Within this limited framework, 
one type of warm inflation solution
was obtained \cite{Berera:1998px}. 
The high-T regime was examined first, since considerable
methodology was already available for treating it. 
However, intrinsically, the statistical state relevant for
warm inflation is not required to be an equilibrium
state.   The slowly varying nature of  the macroscopic variables
in warm inflation cosmology suggest that the statistical state
may not be far from equilibrium, although this is something
that should be proven from the dynamics.  Much work remains
in order to develop the mathematical formalism necessary to
address this problem.
As one step in this direction to fill the missing gaps, we
studied the zero temperature dissipative
dynamics of interacting scalar field 
systems in Minkowski spacetime \cite{Berera:2001gs} 
(for another interesting
direction see \cite{Lawrie:2002wm}).
This is useful to understand, since
the zero temperature limit constitutes a baseline effect, that
will be prevalent in any general statistical state.
The key result presented in this talk is that
for a broad range of cases, involving interaction
with as few as one or two fields, we find dissipative regimes
for the scalar field system.  This is important for inflationary
cosmology, since it suggests dissipation
may be the norm not exception for an interacting scalar field system,
and so warm
inflation is a natural dynamics once interactions are properly 
treated.

A key mechanism we identified which leads to robust warm inflation
involved the scalar inflaton field $\phi$ exciting a heavy
bosonic field $\chi$ which then decays to light fermions
$\psi_d$ \cite{Berera:2001gs},
\begin{equation}
\phi \rightarrow \chi \rightarrow \psi_d.
\label{wimech}
\end{equation}
Recently we studied the simplest model that yields
this mechanism \cite{Berera:2003kg},
\begin{equation}
{\cal L}_I = - \frac{1}{2}g^2 \phi^2 \chi^2 - 
g' \phi {\bar \psi_{\chi}} \psi_{\chi} - h \chi {\bar \psi_d}\psi_d ,
\label{lint}
\end{equation}
where $\psi_d$ are the light fermions to which $\chi$-particles
can decay, with 
\begin{equation}
m_\chi > 2m_{\psi_d}.
\end{equation}
Aside from the last term in Eq. (\ref{lint}), 
this is the typical Lagrangian used in
studies of reheating after inflation 
\cite{Kofman:1997yn,Greene:1998nh,Finelli:2000ya}.
Later we will briefly discuss that in minimal
SUSY extensions of the typical reheating model, decay channels for
the $\chi$ or $\psi_{\chi}$ particles 
are present and the $\psi_d$ field above
is simply a representative example.
Since in the moderate to strong perturbative regime, reheating
models will require SUSY for controlling radiative corrections,
Eq. (\ref{lint}) with inclusion of the $\psi_d$ field thus
is a toy model representative of the typical reheating model.

To study inflation, the effective
evolution equation must be derived
for the background component of the inflaton,
$\varphi \equiv \langle \phi \rangle$.  
The conventional approach
\cite{Guth:1980zm,Sato:ds,Albrecht:1982wi,Linde:1981mu,Copeland:1994vg,Linde:gd,Kofman:1997yn,Greene:1998nh,Finelli:2000ya,Linde:1993cn},
assumes
that aside from radiative corrections that
modify the $\varphi$-effective potential, $V_{\rm eff}(\varphi)$,
this equation is
the same as its classical counterpart.
However, we have shown in earlier works 
\cite{Berera:1998gx,Berera:2001gs}, 
that in addition to
radiative corrections, quantum effects also arise
in the $\varphi$-effective EOM in terms
of temporally nonlocal terms,
which generically lead to
dissipative effects.  
Moreover, although SUSY can cancel large
quantum effects in the local limit, it can not cancel for the dynamical
problem the nonlocal quantum effects.

Here results are presented from \cite{Berera:2003kg},
where we have extended the calculation
to the expanding case (for related earlier works
see \cite{Ringwald:1987ui,Morikawa:dz,Lawrie:xu}).   
Also in \cite{Berera:2003kg} extensive
numerical analysis of dissipative effects was done, which up to
now have only been examined in simplified analytic approximations.
The $\varphi$-effective EOM from \cite{Berera:2003kg} 
for model Eq. (\ref{lint}) is,
\begin{eqnarray}
{\ddot \varphi}(t) & + & 3H(t) {\dot \varphi}(t) + 
\xi {\rm R}(t) \varphi(t) + \frac{dV_{\rm eff}(\varphi(t))}{d\varphi(t)}
 +  g^4 \varphi(t) \int_{t_0}^t dt' \varphi(t') {\dot \varphi}(t') K(t,t')
=0,
\label{eom}
\end{eqnarray}
where
\begin{eqnarray}
K(t,t') & = & \int_{t_0}^{t'} dt'' \int \frac{d^3q}{(2\pi)^3}
\sin \left[2\int_{t''}^t d\tau \omega_{\chi}(\tau)\right] 
\frac{\exp\left[-2
\int_{t''}^t d\tau \Gamma_{\chi}(\tau)\right]}
{4\omega_{\chi}(t) \omega_{\chi}(t'')} ,
\label{kernel}
\end{eqnarray}
\begin{eqnarray}
\omega_{\chi}(\tau) = \left[{\bf q}^2 \frac{a^2(t)}{a^2(\tau)} 
+ m_{\chi}^2(\tau) +
2(6\xi -1)H^2 \right]^{1/2},
\label{omega}
\end{eqnarray}
$m_{\chi}=g\varphi \gg 2m_{\psi_d}$,
$R$ curvature scalar defined below Eq. (\ref{amapprox}),
$\xi$ the gravitational coupling,
$a(t)=\exp(Ht)$ the scalar factor,
$H=\sqrt{8 \pi V_{\rm eff}/(3m_p^2})$ the Hubble parameter, and
\begin{equation}
\Gamma_{\chi}(t) = \frac{h^2 m_{\chi}^2}{8\pi\omega_{\chi}(t)}
\left(1-\frac{4m_{\psi_d}^2}{m_{\chi}^2}\right)^{3/2}
\end{equation}
the decay width.
The kernel $K(t,t')$ Eq. (\ref{kernel}) is obtained by a
linear response approximation, equivalent to the closed time 
path formalism at leading nontrivial order, which treats the
effect of the field $\chi$ on the evolution of $\varphi(t)$.
In the limit $H \rightarrow 0$,
$a \rightarrow constant$, Eq. (\ref{kernel})
agrees with the corresponding kernel for
nonexpanding spacetime in \cite{Berera:2001gs}.
The physical origin of the nonlocal (dissipative) 
term in Eq. (\ref{eom}) is as
follows.  The evolving background field $\varphi$ changes the mass
of the $\chi$ boson which results in the mixing of
its positive and negative frequency modes.  This in turn
leads to coherent production of $\chi$ particles, which then decohere 
through decay into the lighter $\psi_d$-fermions.
A version of the above calculation in nonexpanding spacetime
based on a canonical approach
was done in \cite{Berera:2001gs,Morikawa:dz}.
This approach exhibits much more clearly 
the above picture relating
$\varphi$-dissipation
and particle creation.

The $\psi_{\chi}$-fermions in Eq. (\ref{lint}) have played no role in
the dissipative effects.  Within this toy model,
these fermions are used to mimic SUSY by canceling the quantum
corrections from the $\chi$-boson sector.
Consider firrt Minkowski spacetime.  The one-loop contributions to
the effective potential are from the $\chi$-loop
\begin{equation}
V_1(\varphi) =\frac{1}{2} \int \frac{d^3k}{(2\pi)^3}
E_{m_{\chi}},
\label{vc}
\end{equation}
where
$E_{m_{\chi}}  =  \sqrt{{\bf k}^2 + g^2 \varphi^2}$.
and from the $\psi_{\chi}$-loop
\begin{equation}
V_1(\varphi) =-2 \int \frac{d^3k}{(2\pi)^3}
E_{m_{\psi_{\chi}}},
\label{vpc}
\end{equation}
where setting $g'=g$ in Eq. (\ref{lint}),
$E_{m_{\psi_{\chi}}}  =  \sqrt{{\bf k}^2 +  g^2 \varphi^2}$.
Here the fields have zero explicit mass terms, so 
$m_{\chi}=m_{\psi_{\chi}} = g \varphi$.
Thus if we slightly modify our model Eq. (\ref{lint}) so that
there are four $\chi$ fields
for the one $\psi_{\chi}$-fermion,
then summing Eqs. (\ref{vc})+(\ref{vpc}), these contributions
cancel., leaving $V_{eff}(\varphi)$ to be simply whatever
is chosen for the $\varphi$-potential.
Moreover, since the $\varphi$-potential must be very flat,
thus weakly coupled, quantum corrections arising from
the $\phi$-field are negligible. 
To extend to de Sitter spacetime,
these results are modified due to $\xi$-dependent
mass corrections and the $k_0$ integration in this geometry.
Both modifications add corrections from the Minkowski spacetime
effective potential by terms
$\sim O(\ll 1) g^2 H^2 \varphi^2  < m^2_{\phi} \varphi^2$
\cite{Vilenkin:sg}
and so can be neglected.

When the motion of $\varphi$ is slow,
which is the regime of
interest for inflation, a adiabatic-Markovian approximation
can be applied that converts Eq. (\ref{eom}) to one that is completely
local in time, albeit with time derivative terms.
The Markovian approximation amounts to substituting
$t' \rightarrow t$ in the arguments of the $\varphi$-fields
in the nonlocal term in Eq. (\ref{eom}).
The adiabatic approximation then requires that all macroscopic
motion is slow on the scale of microscopic motion, thus
\begin{eqnarray}
\frac{\dot \varphi}{\varphi} & < & H, \Gamma_{\chi}
\nonumber \\
H & < & \Gamma_{\chi}.
\end{eqnarray}
Moreover when also $H < m_\chi$,
the kernel $K(t,t')$ above is well approximated by the nonexpanding
limit $H \rightarrow 0$.  Combining both these
approximations, the effective EOM Eq. (\ref{eom}) takes on
the form in Eq. (\ref{amapprox}),
where, by defining $\alpha_\chi = h^2 /(8 \pi )$,
                                                                                
\begin{equation}
\Upsilon(\varphi) = \frac{\sqrt{2} g^4 \alpha_\chi \varphi^2 }
{64\pi m_\chi \sqrt{1 + \alpha^2_{\chi}}
\sqrt{\sqrt{1 + \alpha^2_{\chi}}+1}} .
\end{equation}

\FIGURE{\epsfig{file=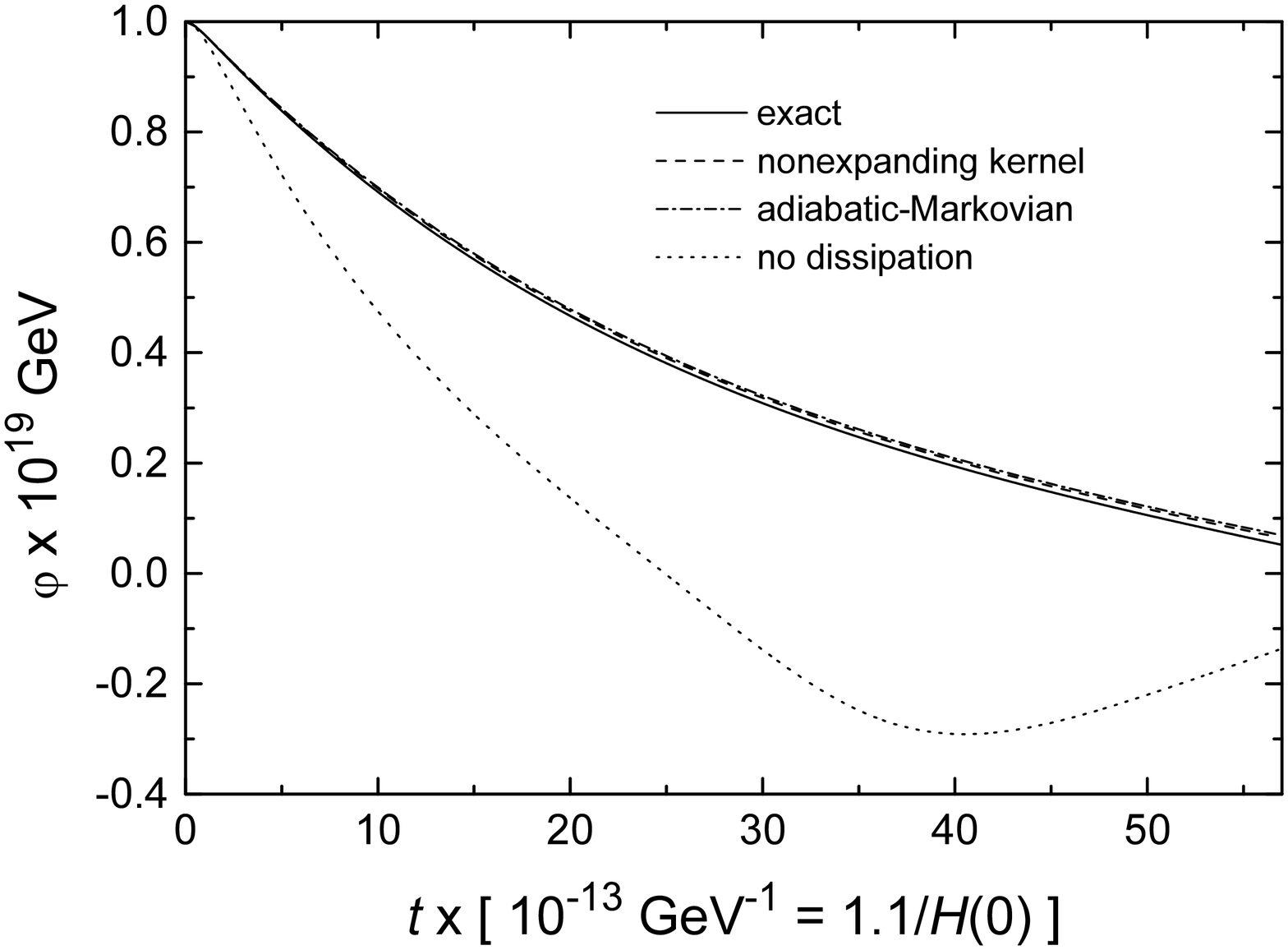,angle=0,width=8.7cm}
\caption{Evolution of $\varphi(t)$ for $\lambda=10^{-13}$,
$g=h=0.37$, $\xi=0$,
$\varphi(0) = m_p$, ${\dot \varphi}(0)=0$.}
        \label{fig1}}

Fig. \ref{fig1} compares the various approximations 
for a representative case, where $g=h=0.37$ and the inflaton potential
is that for a chaotic inflation
$V_{\rm eff}(\varphi) = \lambda \varphi^4/4$
with $\lambda = 10^{-13}$ \cite{Linde:gd}.
In Fig. \ref{fig1}  evolution has been examined
at the final stages of chaotic inflation, where we start with
$\varphi(t_0=0) = m_p$.  The solid line is the exact result based on
numerically solving Eq. (\ref{eom}).
Plotted alongside this, although almost
indiscernible, is the same solution expect using
the nonexpanding spacetime kernel (dashed line),
obtained by setting $H \rightarrow 0$, $a \rightarrow constant$
in Eq. (\ref{kernel}), and the solution based on
the adiabatic-Markovian approximation of Eq. (\ref{amapprox})
(dot-dashed line) for the same parameter set.
As seen, the expanding and nonexpanding cases differ
by very little and the adiabatic-Markovian approximation is in
good agreement with the exact solution.
This confirms simplifying approximations
claimed in \cite{Berera:1998gx,Berera:2001gs,Ringwald:1987ui,Morikawa:dz}
but up to now had not been numerically verified.

More interestingly, and the first major result in \cite{Berera:2003kg},
the dotted line in Fig. \ref{fig1} is the solution that would be found
by the conventional approach in which the nonlocal term
in Eq. (\ref{eom}) is ignored.
By the conventional approach 
\cite{Albrecht:1982wi,Linde:1981mu,Linde:gd,Kofman:1997yn,Greene:1998nh,Finelli:2000ya,Copeland:1994vg,Linde:1993cn},
one expects
the inflaton to start oscillating, which is the precursor
to entering various stages of pre/re-heating.
However with account for dissipative effects, this
never happens, since the inflaton remains 
overdamped till the end when it settles at its minima at $\varphi=0$.
Moreover, throughout inflation, and not just in the final stages,
the inflaton
is dissipating
energy and yielding a radiation component of magnitude
\begin{equation}
\rho_r \approx \frac{\Upsilon {\dot \varphi}^2}{4H} .
\label{rad}
\end{equation}

For the case in Fig. \ref{fig1}, 
the overdamped regime for the inflaton persists
until $g \lesssim 0.35$, below which its evolution at the end
of inflation has oscillatory features similar to conventional
expectations.  However
the inflaton is still dissipating radiation 
at the level Eq. (\ref{rad}) all throughout the inflation period.
If the temperature scale associated with the radiation
energy density produced through this dissipative
mechanism is greater than the inflaton mass, $T > m_{\phi}$, then
inflaton fluctuations will be significantly altered from their zero temperature
behavior.  Moreover if $\Gamma_{\chi} > H$, it is possible for
the radiation to thermalize, in which case the inflaton fluctuations
will now be thermal.  For cold inflation models, it is generally
necessary that $m_{\phi} \stackrel{<}{\sim} H$.  Thus for all cold inflation
models, the basic criteria is that if $\rho_r > H^4$, then the radiation
has important observational effects in terms of its influence on
inflaton fluctuations.  For our model Eq. (\ref{lint}),
$\rho_r > H^4$
and $\Gamma_{\chi} > H$,
if for example  $g=h$ and $g > 10^{-2}$ or
as another example if $h=0.1$ and
$g > 10^{-3}$.
Thus fairly small couplings already lead to observationally significant thermal
fluctuations in the inflaton.
We thus arrive at the second major result in \cite{Berera:2003kg}.  
In multifield inflation models, or 
any inflation model once reheating interactions are accounted for,
there are parameter regimes feasible to inflation,
that have never been properly treated since
the nonlocal term in Eq. (\ref{eom}) is neglected.
Once this term is retained,
it is seen that up to fairly weak coupling,
adequate radiation is produced during inflation to alter
density perturbations.
Although this conclusion is
based on situations where thermalization is assumed,
one can infer the same qualitative
outcome once $\rho_r > H^4$,
irrespective of the statistical state.

Underlying the results here on energy dissipation and inflaton
evolution damping is the interaction scheme Eq. (\ref{wimech}).
Such an interaction scheme is very common in particle physics models.
For example, the simplest implementation of
the Higgs mechanism in the Standard Model has the background Higgs
field coupled to W and Z bosons, thereby generating their masses,
and these bosons then are coupled to light fermions through
the well known charged and neutral current interactions.
Also in simple SUSY models, the interaction scheme Eq. (\ref{wimech})
is very common.  For example
a minimal SUSY extension that incorporates the $\phi-\chi$ coupling
would be
\begin{equation}
W= \sqrt{\lambda} \Phi^3 + g \Phi X^2 + f X^3 + m X^2,
\label{susymodel}
\end{equation}
where $\Phi = \phi + \psi \theta + \theta^2 F$ and 
$X = \chi+ \theta \psi_{\chi} + \theta^2 F_{\chi}$
are chiral superfields.  In the above model, there would be
no additional fermion to associate with $\psi_d$ from our toy
model Eq. (\ref{lint}).  However the $\chi$-field has a decay channel
via a $\psi_{\chi}$ triangle-loop into two light inflaton bosons $\phi$.
For this case, everything in Eqs. (\ref{eom})-(\ref{omega}) 
is unaltered except
the decay channel is different with now 
$\Gamma_{\chi} \sim (fg^2)^2 m_{\chi}$.
Thus there are additional factors of coupling constants,
but at moderate coupling, radiation production would be sufficiently large
to affect density perturbations and at the very largest perturbative regime
the evolution of $\varphi$ could be altered into the overdamped behavior
similar to the solid line in Fig. \ref{fig1}.  

In general for large interaction couplings, such as
$g \stackrel{>}{\sim} 10^{-4}$ for the $\phi-\chi$ interaction in our
model Eq. (\ref{lint}), to maintain the flat potential $V_{eff}(\varphi)$,
SUSY is needed.  This situation arises in our warm inflation model
as well as in many cold inflation models such as hybrid
inflation \cite{Copeland:1994vg,Linde:1993cn} 
or any cold inflation model where interactions to
fields introduced for reheating are large,
such as in
\cite{Kofman:1997yn,Greene:1998nh,Finelli:2000ya}.
Thus generic SUSY
extensions to common cold inflation models readily have
interaction structures of the form Eq. (\ref{wimech}), and
so the resulting dynamics departs radically from 
the cold inflation picture that is tacitly assumed.

\section{Observational tests}
\label{sect3}

As stated in the last section, when a thermalized radiation component
is present with $T > m_{\phi}$, inflaton fluctuations are dominantly
thermal rather than quantum.  There are two distinct regimes of
warm inflation to note.  One is the weak dissipative 
regime \cite{Moss:wn,Berera:1995wh},
\begin{equation}
\delta \varphi^2  \sim HT \hspace{0.2cm}
{\rm warm \hspace{0.1cm} inflation}
\hspace{0.1cm} (\Upsilon < 3H), \hspace{0.2cm} T > m_{\phi},
\end{equation}
and the other is the strong dissipative regime \cite{Berera:1999ws}, 
\begin{equation}
\delta \varphi^2 \sim \sqrt{H\Upsilon}T \hspace{0.2cm}
{\rm warm \hspace{0.1cm} inflation} \hspace{0.1cm}(\Upsilon > 3H),
\hspace{0.2cm} T > m_{\phi}.
\label{dpsd}
\end{equation}
For comparison, for cold inflation, where inflaton
fluctuations are exclusively quantum \cite{Guth:ec}, 
\begin{equation}
\delta \varphi^2 \sim H^2, \hspace{0.2cm}
{\rm cold \hspace{0.1cm}inflation} \hspace{0.2cm} T < m_{\phi} .
\end{equation}
For both cold and warm inflation, density perturbations are obtained by
the same expression,
$\delta \rho/\rho \sim H \delta \varphi/{\dot \varphi}$.

In \cite{Taylor:2000ze} an order of magnitude 
estimate of density perturbations
during warm inflation was computed by matching the thermally
produced fluctuations to gauge invariant parameters when the fluctuations
cross the horizon (for other phenomenological treatments of warm inflation see
\cite{Lee:1999iv,Bellini:1999ag,DeOliveira:2001he,Chimento:2002us,Hwang:2001fb,Lee:2003ed}).
This work provided a clear statement of the consistency
condition.
Cold inflation has three parameters,  
related to the potential energy
magnitude $V_0$, slope
$\epsilon = m^2_{pl} V'/(16 \pi V)$, 
and curvature $\eta$ Eq. (\ref{etadef}), whereas
there are four observable constraints ($\delta_H$, 
$A_g$, $n_s$, $n_g$).
This implies a redundancy in the observations and
allows for a consistency relation \cite{Liddle:1993fq}. This is usually 
expressed as a relationship
between the tensor-to-scalar ratio and the slope of the tensor spectrum.
Warm inflation has an extra parameter, the dissipation factor,
which implies four constraints for four parameters. Hence we do not expect the
consistency relation of standard inflation 
to hold in warm inflation \cite{Taylor:2000ze}.
Thus discriminating between warm and standard 
inflation requires measuring
all four observables. The WMAP and upcoming Planck satellite missions
should provide strong constraints on the scalar spectrum
and having polarization detectors, it is hoped
the tensor spectrum also will be measured.
At the same level of approximation,
nongaussian effects from warm inflation models
were computed and found to be of the same
order of magnitude as in the cold inflation case, and
thus too small to be measured \cite{Gupta:2002kn,Gupta:2003au}.

More accurate treatments of density perturbations have followed 
\cite{Lee:1999iv,DeOliveira:2001he,Chimento:2002us,Hwang:2001fb,Lee:2003ed},
which use the cosmological perturbation equations.
Following our recent results \cite{Hall:2003zp}, working in the zero-shear
gauge with perturbed metric 
\begin{equation}
ds^2 = -(1-2\varsigma)dt^2 + a^2(1+2\varsigma)\delta_{ij}dx^i dx^j,
\end{equation}
we numerically evolved
the Einstein and scalar field perturbation equations
\begin{eqnarray}
{\dot \varsigma} + H \varsigma + 4\pi G k^{-1}a(p+\rho)v = 0,
\nonumber \\
3H {\dot \varsigma} + (3H^2 + k^2a^{-2})\varsigma - 4 \pi G \delta \rho =0,
\nonumber \\
{\ddot \varsigma} + 4H {\dot \varsigma} + (2{\dot H}+3H^2) \varsigma
+ 4 \pi G \delta p =0, \\
\delta {\ddot \varphi} + (3H + \Upsilon){\dot \varphi} + 
{\dot \varphi} \delta \Upsilon + k^2a^{-2} \delta \varphi 
+ \delta V_{,\varphi} + 4 {\dot \varphi}{\dot \varsigma}
& - & \Upsilon {\dot \varphi} \varsigma - 2V_{,\varphi}\varsigma =0,
\end{eqnarray}
where
\begin{eqnarray}
\delta \rho = {\dot \varphi} \delta {\dot \varphi} +
V_{,\varphi}\delta \varphi + {\dot \varphi}^2 \varsigma + T \delta s,
\nonumber \\ 
\delta p = {\dot \varphi} \delta {\dot \varphi} -
V_{,\varphi}\delta \varphi + {\dot \varphi}^2 \varsigma + s \delta T.
\end{eqnarray}
Here $\delta \varphi$ is the inflaton perturbation, $\delta T$ is the 
temperature fluctuation, $v$ is the velocity perturbation, and
$s$ is the entropy density.

As a example we examined the generic symmetry breaking potential
\begin{equation}
V = \frac{1}{4} \lambda (\phi^2-\phi^2_0)^2.
\label{sympot}
\end{equation}
One of the most interesting outcomes of our study was that by accounting
for dissipative effects, the spectral index generically runs with wavenumber.
As one interesting example, if we consider a model where
$\Upsilon \sim c \varphi^2$ in Eq. (\ref{amapprox}),
we obtain an index that runs from
blue at large scales to red at small scales, such as 
in Fig. \ref{fig2}.
This is an interesting result for the current observational situation.
The WMAP CMB first year data suggests a spectral index 
$n_s < 1$ \cite{Bennett:2003bz}.
However, when this data is taken in combination with large scale structure
data \cite{Peiris:2003ff}, 
the index then has a form similar to Fig. \ref{fig2}.
Also in \cite{Hall:2003zp} the effect
of dissipation and temperature for the model Eq. (\ref{sympot})
is shown can yield an oscillatory power spectrum, which is rare
to most models, thus could be a robust signature
for warm inflation.

\FIGURE{\epsfig{file=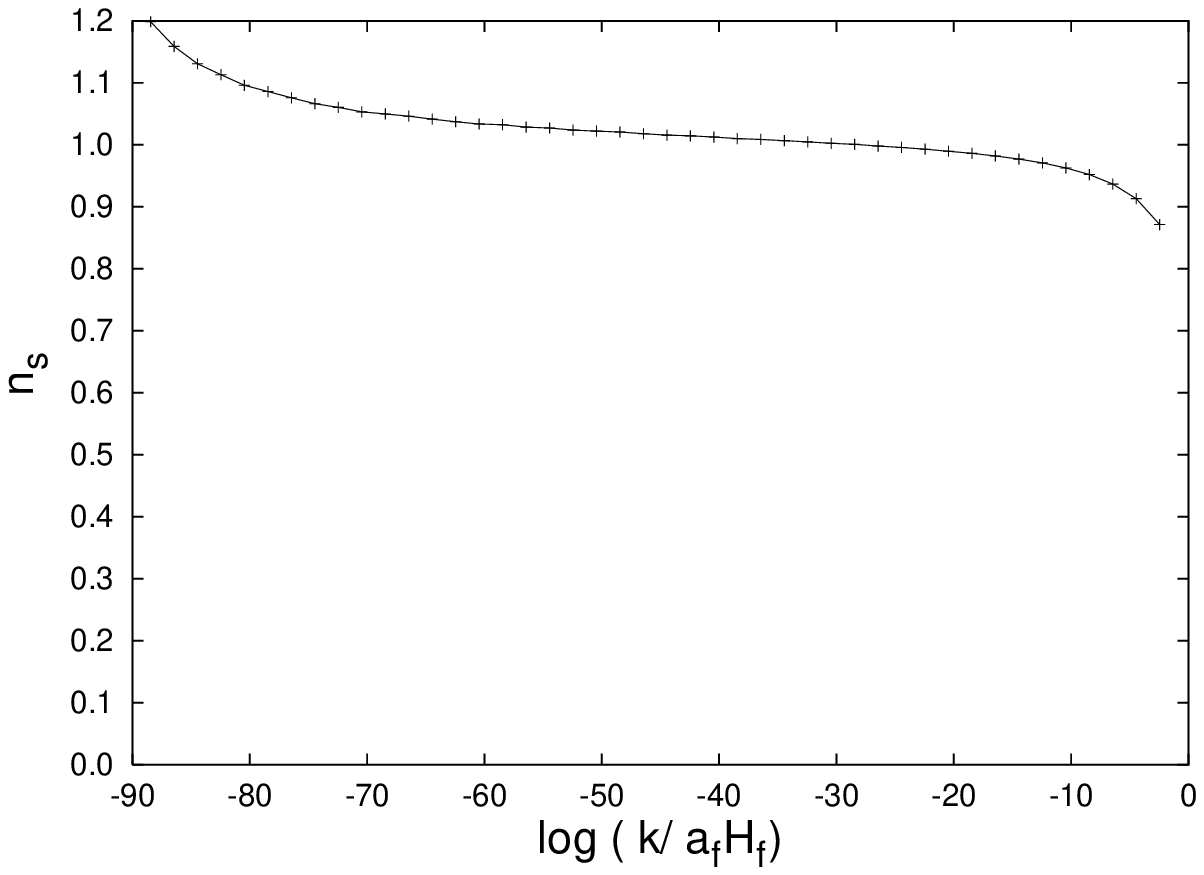,angle=0,width=8.7cm}
\caption{The scalar spectral index for the potential Eq. (\ref{sympot})
with damping term $\Upsilon \sim \varphi^2$.
The wavenumber has been normalized by the horizon size at the
end of inflation.}
        \label{fig2}}

\section{Solution to eta-problem}
\label{sect4}

The warm inflation solution is
examined for the simple potential
\begin{equation}
V = \frac{1}{2} m_{\phi}^2 \phi^2.
\label{quadpot}
\end{equation}
In the cold inflation case, such a model requires an initial
inflaton amplitude $\langle \phi \rangle = \varphi > m_{pl}$.
Moreover, SUSY models that realize a potential like this
inevitably lead to an eta-problem based on the reasons
discussed in the Introduction.  

Let us now treat this model in the warm inflation case.
To focus on the essential
points, our calculations here will be purely phenomenological,
although they can be readily derived from a first principles
quantum field theory calculation.  We consider the case
where the dissipative coefficient in Eq. (\ref{amapprox})
is independent of both $\varphi$ and $T$, $\Upsilon = constant$.
The inflaton initially is at a nonzero field
amplitude $\varphi \ne 0$, thus supporting a vacuum energy. 

The background cosmology for models with constant $\Upsilon$
and monomial potentials has been solved exactly \cite{Berera:1996fm}.
From this we find $N_e \approx H\Upsilon/m_{\phi}^2$.
The radiation production is determined from the energy conservation
equation, 
\begin{equation}
{\dot \rho}_r + 4H \rho_r = \Upsilon {\dot \varphi}^2.
\label{econs}
\end{equation}
During warm inflation ${\dot \rho}_r \approx 0$ 
\cite{Berera:1995ie,Berera:1995wh}, so that
Eq. (\ref{econs}) reduces to Eq. (\ref{rad}).
Identifying $\rho_r \sim T^4$ permits determination of 
the temperature during warm inflation.
Finally, once $T$ is determined, Eq. (\ref{dpsd}) allows
determination of density perturbations.

Combining these expressions, for model Eq. (\ref{quadpot}) with
$\Upsilon = const.$ in Eq. (\ref{amapprox}), gives
\begin{equation}
N_e \approx 2\sqrt{2} \frac{\Upsilon \varphi_0}{m_{\phi} m_{pl}}
\end{equation}
\begin{equation}
T \approx \frac{m_{\phi}^{3/4} m_{pl}^{1/4} \varphi_0^{1/4}}{\Upsilon^{1/4}}
\end{equation}
\begin{equation}
\frac{\delta \rho}{\rho} \approx \left(\frac{\varphi_0}{m_{\phi}}\right)^{3/8}
\left(\frac{\Upsilon}{m_{pl}}\right)^{9/8}.
\end{equation}
Imposing observational constraints $N_e =60$ and
$\delta \rho/\rho = 10^{-5}$, leads to the results
\begin{equation}
\frac{m_{\phi}}{H} \approx 5.5 \times 10^{-9} \frac{m_{pl}}{m_{\phi}},
\end{equation}
\begin{equation}
\frac{\varphi_0}{m_{pl}} \approx 5.3 \times 10^{8} \frac{m_{\phi}}{m_{pl}},
\end{equation}
$\Upsilon \approx 4 \times 10^{-8} m_{pl}$,
and $T \approx 10^4 m_{\phi}$, with the ratio $m_{\phi}/m_{pl}$
free to set.  For $m_{\phi}/m_{pl} \stackrel{<}{\sim} 10^{-9}$, it means
$\eta > 1$ and $\varphi < m_{pl}$.
Thus we see for sufficiently small
inflaton mass, $m_{\phi} \stackrel{<}{\sim} 10^{10}{\rm GeV}$,
there is no eta-problem, since $m_{\phi}\gg H$
and $\varphi < m_{pl}$.  Since this warm 
inflation solution works for $\eta \gg 1$, SUSY models
realizing simple monomial potentials like Eq. (\ref{quadpot})
do not require any special symmetries, as is the case
discussed in the Introduction for cold inflation models.
The "eta" and large $\varphi$-amplitude problems simply
correct themselves once interactions already present in the
models are properly treated.

\section{Conclusion}
\label{sect5}

Warm inflation is seen to have several remarkable
features.   In particular, warm inflation 
\begin{itemize}

\item is generic in quantum field theory

\item dissipative effects can produce a running scalar spectral index

\item has no eta-problem: $m_{\phi} \gg H$ ($\eta \gg 1$) is permissible

\item has no large $\varphi$ amplitude problem: $\varphi < m_{pl}$

\item has no graceful exit problem: inflation automatically terminates
into a radiation dominated regime

\item has no quantum-to-classical issues: inflaton fluctuations
are classical upon inception.

\end{itemize}

The dissipative effects discussed in Sec. \ref{sect2} also can serve to damp
kinetic energy contributions before inflation, thus alleviating the initial
condition problem of inflation
\cite{Berera:2000xz}.
Warm inflation is also a conducive regime for the creation of large scale
cosmic magnetic fields based on the ferromagnetic Savvidy vacuum
scenario
\cite{Berera:1998hv}.
Also, spontaneous baryogenesis has been shown to work efficiently
in the last stages of warm inflation
\cite{Brandenberger:2003kc}.

Progress toward a theory of inflation requires completely
understanding the dynamics of inflation models.
Our work, more correctly quantum field theory, demonstrates
that inflation models generically are dissipative systems.
These effects crucially influence inflaton evolution and observational 
signatures such as density perturbations.  Moreover, dissipative effects
play a central role in eliminating the
eta and the large $\varphi$-amplitude problems.

\end{document}